\documentclass[11pt]{llncs}
\usepackage{amsmath}
\usepackage{amssymb}
\usepackage{path}
\usepackage{xspace}
\usepackage{graphicx}
\usepackage[font=small, format=hang, labelfont=bf]{caption}
\usepackage[subrefformat=parens, labelfont=default]{subfig}
\usepackage{fullpage}
\usepackage{hyperref}
\usepackage{algorithm}
\usepackage{algorithmic}

\title{Approximating Minimum Manhattan Networks in Higher Dimensions}

\author{
  Aparna~Das\inst{1}
  \and
  Emden~R.~Gansner\inst{2}
  \and 
  Michael~Kaufmann\inst{3}
  \and 
  Stephen~Kobourov\inst{1}
  \and 
  Joachim~Spoerhase\inst{4}
  \and 
  Alexander~Wolff\inst{4}
}

\authorrunning{A.~Das et al.}

\institute{
  Dept.~of Comp.~Sci., University of Arizona, Tucson, AZ, U.S.A.
  \and
  AT\&T Labs Research, Florham Park, NJ, U.S.A.
  \and
  Wilhelm-Schickard-Institut f\"ur Informatik, Universit\"at
  T\"ubingen, Germany 
  \and
  Institut f\"ur Informatik, Universit\"at W\"urzburg, Germany
}

\newcommand{\R}{\ensuremath{\mathbb{R}}}

\newcommand{\NE}{\ensuremath{\mathrm{NE}}\xspace}

\newcommand{\SW}{\ensuremath{\mathrm{SW}}\xspace}
\newcommand{\MSN}{\ensuremath{\pi_\mathrm{SN}}\xspace}
\newcommand{\MWE}{\ensuremath{\pi_\mathrm{WE}}\xspace}
\newcommand{\tE}{\ensuremath{t_\mathrm{E}}\xspace}
\newcommand{\tW}{\ensuremath{t_\mathrm{W}}\xspace}
\newcommand{\tS}{\ensuremath{t_\mathrm{S}}\xspace}
\newcommand{\tN}{\ensuremath{t_\mathrm{N}}\xspace}
\newcommand{\Nopt}{\ensuremath{N_\mathrm{opt}}\xspace}
\newcommand{\Nhor}{\ensuremath{N^\mathrm{hor}}\xspace}
\newcommand{\Nver}{\ensuremath{N^\mathrm{ver}}\xspace}
\newcommand{\Mver}{\ensuremath{M^\mathrm{ver}}\xspace}
\newcommand{\cost}{\ensuremath{\mathrm{weight}}\xspace}
\newcommand{\dir}{\ensuremath{\mathrm{dir}}}
\newcommand{\Nopthor}{\ensuremath{\Nopt^\mathrm{hor}}\xspace}
\newcommand{\Noptver}{\ensuremath{\Nopt^\mathrm{ver}}\xspace}
\newcommand{\Nxy}{\ensuremath{N_{xy}}\xspace}
\newcommand{\Txy}{\ensuremath{T_{xy}}\xspace}
\newcommand{\eps}{\ensuremath{\varepsilon}\xspace}
\newcommand{\opt}{\ensuremath{\mathrm{OPT}}\xspace}

\newcommand{\optup}{\ensuremath{\opt^\mathrm{up}}\xspace}

\newcommand{\red}{\ensuremath{R}\xspace}
\newcommand{\blue}{\ensuremath{B}\xspace}
\newcommand{\BRP}{\ensuremath{\mathrm{BRP}}\xspace}
\newcommand{\rpatch}{\ensuremath{r_\mathrm{patch}}}
\newcommand{\etal}{et al.}

\graphicspath{{pic/}}

\newtheorem{observation}{Observation}
\newenvironment{pf}{\begin{proof}}{\qed\end{proof}}

\begin{document}

\maketitle

\begin{abstract}
  We study the minimum Manhattan network problem,
  which is defined as follows.  Given a set of points called
  \emph{terminals} in~$\R^d$, find a minimum-length network such that
  each pair of terminals is connected by a set of axis-parallel line
  segments whose total length is equal to the pair's Manhattan (that
  is, $L_1$-) distance.  The problem is NP-hard in 2D and there is no
  PTAS for 3D (unless ${\cal P}\!=\!{\cal NP}$). Approximation
  algorithms are known for 2D, but not for 3D.
  
  We present, for any fixed dimension~$d$ and any $\eps>0$, an
  $O(n^\eps)$-approxi\-ma\-tion algorithm. For 3D, we also give a
  $4(k-1)$-approximation algorithm for the case that the terminals are
  contained in the union of $k \ge 2$ parallel planes.
\end{abstract}

\section{Introduction}

In a typical network construction problem, one is given a set of
objects to be interconnected 
such that some constraints regarding the connections are fulfilled. 
Additionally, the network must be of little cost.  
For example, if the objects are points in Euclidean
space and the constraints say that, for some fixed $t>1$, each pair of
points must be connected by a path whose length is bounded by~$t$
times the Euclidean distance of the points, then the solution is a
so-called \emph{Euclidean $t$-spanner}.  Concerning cost, one usually
requires that the total length of the network is proportional to the
length of a Euclidean minimum spanning tree of the points.  Such low-cost
spanners can be constructed efficiently~\cite{admss-esstl-95}.

In this paper, we are interested 
in constructing 1-spanners, with respect to the
Manhattan (or $L_1$-) metric.  Rather than requiring that the total
length of the network 
is proportional to the minimum spanning tree of the points, 
our aim is to minimize the total length
(or \emph{weight}) of the network.  Note that the Euclidean 1-spanner
of a set of points is simply the complete graph (if no three points
are collinear) and hence, its weight is completely determined.
Manhattan 1-spanners, in contrast, have many
degrees of freedom and vastly different weights.

\begin{figure}[tb]
  \subfloat[an M-network for $T=\{s,s',t,t'\}$\label{sfg:2D-MN}]%
  {\parbox{.2\textwidth}{\centering\includegraphics[page=1]{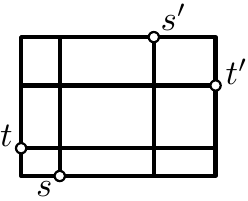}}}
  \hfill
  \subfloat[a minimum M-network for~$T$\label{sfg:2D-MMN}]%
  {\parbox{.19\textwidth}{\hspace*{-2ex}\centering\includegraphics[page=2]{example-2d}}\hspace*{-2ex}}
  \hfill
  \subfloat[a minimum M-network in 3D\label{sfg:3D-MMN}]%
  {\parbox{.24\textwidth}{\hspace*{-5ex}\includegraphics{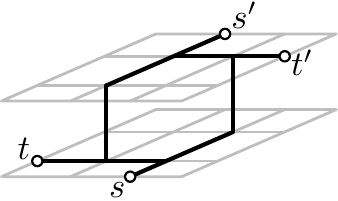}}\hspace*{-5ex}}
  \hfill
  \subfloat[M-paths missing each other\label{sfg:miss}]%
  {\parbox{.2\textwidth}{\includegraphics{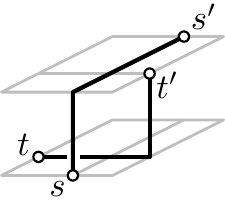}}}
  \caption{Examples of M-networks in 2D and 3D.}
  \label{fig:examples}
\end{figure}

More formally, given two points~$p$ and~$q$ in $d$-dimensional
space~$\R^d$, a \emph{Manhattan path} connecting~$p$ and~$q$ (a
$p$--$q$ \emph{M-path}, for short) is a sequence of axis-parallel
line segments connecting~$p$ and~$q$ whose total length equals the
Manhattan distance between~$p$ and~$q$.  
Thus an M-path is a monotone rectilinear path.
For our purposes, a set of axis-parallel line segments is a
\emph{network}.  Given a network~$N$, its \emph{weight} $\|N\|$ is the
sum over the lengths of its line segments.
A network~$N$ \emph{Manhattan-connects} (or \emph{M-connects}) two
given points~$p$ and~$q$ if it ``contains'' a $p$--$q$ M-path~$\pi$. 
Note that we slightly abuse the notation here: we 
mean pointwise containment, that is, we require 
$\bigcup \pi \subseteq \bigcup N$.
Given a set~$T$ of points---called \emph{terminals}---in~$\R^d$, a
network~$N$ is a \emph{Manhattan network} (or \emph{M-network})
for~$T$ if~$N$ M-connects every pair of terminals in~$T$.  
The \emph{minimum Manhattan network problem} (MMN) consists of
finding, for a given set~$T$ of terminals, a minimum-weight M-network.
For examples, see Fig.~\ref{fig:examples}.

M-networks have important applications in several areas such as VLSI
layout and computational biology. For example, Lam
\etal~\cite{lap-pafst-03} used them in gene alignment in order to 
reduce 
the size of the search space of the Viterbi algorithm for pair hidden
Markov models.

\subsection{Previous work} 

The 2D-version of the problem, 2D-MMN, was introduced by
Gudmundsson \etal~\cite{gln-ammn-01}.  They gave an 8- and a
4-approximation algorithm.  Later, the approximation ratio was
improved to~3~\cite{bwws-mmnpa-06,fs-s3amm-08t} and then to~2, which
is currently the best possible.  It was achieved in three different
ways: via linear programming~\cite{cnv-raamm-08}, using
the primal--dual scheme~\cite{n-eprmc-05} and with purely geometric
arguments~\cite{gsz-yaa2a-08}.  The last two algorithms run in $O(n
\log n)$ time, given a set of~$n$ points in the plane.  A ratio of~1.5 was
claimed~\cite{su-15amm-05}, but apparently the proof is
incomplete~\cite{fs-s3amm-08t}.  
Chin \etal~\cite{cgs-mmnnp-11} finally
settled the complexity of 2D-MMN by proving it NP-hard.

A little earlier, Mu{\~n}oz \etal~\cite{msu-mmnp3-09}
considered 3D-MMN.  They showed that the 
problem is NP-hard and that it is 
NP-hard to approximate beyond a
factor of 1.00002.  
For the special case of 3D-MMN, where any cuboid spanned by two
terminals contains other terminals or is a rectangle, they gave a
$2\alpha$-approximation algorithm, where~$\alpha$ denotes the best approximation
ratio for 2D-MMN.  They posed the design of approximation algorithms
for general 3D-MMN as an open problem.

\subsection{Related problems}

As we observe in Section~\ref{sec:relat-stein-type}, MMN is a special
case of the \emph{directed Steiner forest problem} (DSF).  More
precisely, an instance of MMN can be decomposed into a constant number
of DSF instances.  The input of DSF is an edge-weighted directed
graph~$G$ and a set of vertex pairs.  The goal is to find a
minimum-cost subgraph of~$G$ (not necessarily a forest) that connects
all given vertex pairs.  Recently, Feldman \etal~\cite{fkn-iaadsf-09}
reported, for any $\eps>0$, an $O(n^{4/5+\eps})$-approximation
algorithm for DSF, where~$n$ is the number of vertices of the given
graph.  This bound carries over to $d$D-MMN.

An important special case of DSF is the \emph{directed Steiner
  \emph{tree} problem} (DST).  Here, the input instance specifies an
edge-weighted digraph~$G$, a \emph{root} vertex~$r$, and a subset~$S$
of the vertices of~$G$ to which $r$ must connect.  An optimum solution
for DST is a minimum-weight $r$-rooted subtree of~$G$ spanning~$S$.
DST admits an $O(n^\eps)$-approximation for any~$\eps>0$
\cite{cccdggl-aadsp-98}.

A \emph{geometric} optimization problem that resembles MMN
is the \emph{rectilinear Steiner arborescence problem} (RSA).
Given a set of points in~$\R^d$ with non-negative coordinates, a
rectilinear Steiner arborescence is a spanning tree 
that connects all points with M-paths to the origin.  As in MMN, the
aim is to find a minimum-weight network.  For 2D-RSA, there is a 
polynomial-time approximation scheme (PTAS)
\cite{lr-ptasrsap-00} based on Arora's technique for approximating
geometric optimization problems such as TSP~\cite{a-asnph-03}.   
It is not known whether 2D-MMN admits a PTAS.  
Arora's technique 
does not directly apply here as M-paths between terminals
forbid detours and thus may not respect portals.  

\subsection{Our contribution}

We first present a $4(k-1)$-approximation algorithm for the special
case of 3D-MMN where the given terminals are contained in $k \ge 2$
planes parallel to the $x$--$y$ plane; see Section~\ref{sec:k-planes}.

Our main result is an $O(n^\eps)$-approximation algorithm for
$d$D-MMN, for any $\eps>0$.  We first present the algorithm in detail
for three dimensions; see Section~\ref{sec:general-case}.  Since the
algorithm for arbitrary 
dimensions is a straightforward generalization of the algorithm for 3D
but less intuitive, we describe it in the appendix.

Our $O(n^\eps)$-approximation algorithm for $d$D-MMN constitutes a
significant improvement 
upon the best known ratio of $O(n^{4/5+\eps})$ for (general) directed
Steiner forest \cite{fkn-iaadsf-09}. We obtain this result by
exploiting the geometric structure of the problem.  To underline the
relevance of our result, we remark that the bound of $O(n^\eps)$ is the
best known result also for other directed Steiner-type problems such
as DST~\cite{cccdggl-aadsp-98} or even acyclic
DST~\cite{zelikovsky-adstapprox-97}. 

Our $O(k)$-approximation algorithm for the $k$-planes case relies on
recent work by Soto and Telha~\cite{st-2dorg-11}.
They show that, given a set of red and blue points in the plane, one
can determine efficiently a minimum-cardinality set of points that
together \emph{pierce} all rectangles having a red point in the lower
left corner and a blue point in the upper right corner. Combining this
result with an approximation algorithm for 2D-MMN, 
yields an approximation algorithm for the 2-planes case.  We show how to
generalize this idea to $k$ planes.

\section{Some Basic Observations}\label{sec:basic-observations}

We begin with some notation. Given a point $p \in \R^3$, we
denote the $x$-, $y$- and $z$-coordinate of~$p$ by $x(p)$, $y(p)$, and
$z(p)$, respectively.  Given two points~$a$ and~$c$ in~$\R^2$, let
$R(a,c)=\{ b \in \R^2 \mid x(a) \le x(b) \le x(c), \, y(a) \le y(b)
\le y(c) \}$ be the \emph{rectangle spanned by~$a$ and~$c$}.
If a line segment is parallel to the $x$-, $y$-, or $z$-axis, we say
that it is $x$-, $y$-, or $z$-\emph{aligned}.
In what follows, we consider the
3-dimensional case of the MMN problem, unless otherwise stated.

\subsection{Quadratic Lower Bound for Generating Sets in 3D}
\label{sec:generating}

Intuitively, what makes 3D-MMN more difficult than 2D-MMN is the following: 
in 2D, if the bounding box of terminals~$s$ and~$s'$ and the
bounding box of~$t$ and~$t'$ cross (as in Fig.~\ref{sfg:2D-MMN}), then
any $s$--$s'$ M-path will intersect any $t$--$t'$ M-path, which yields
$s$--$t'$ and $t$--$s'$ M-paths for free (if~$s$ and~$t$ are the lower
left corners of their respective boxes).  A similar statement for 3D
does not hold; M-paths can ``miss'' each other---even if their
bounding cuboids cross; see Fig.~\ref{sfg:miss}.

Let us formalize this observation.  Given a set~$T$ of
terminals, a set~$Z$ of pairs of terminals is a \emph{generating
  set}~\cite{kia-iammn-02} if any network that M-connects the pairs
in~$Z$ in fact M-connects \emph{all} pairs of terminals.
In~2D, any MMN instance has a generating set of linear
size~\cite{kia-iammn-02}.  Unfortunately this
result does not extend to~3D.  Below, we construct an
instance that requires a generating set of size $\Omega(n^2)$.
The idea of using linear-size generating sets is 
exploited by several algorithms for
2D-MMN~\cite{cnv-raamm-08,kia-iammn-02}. The following theorem shows that these 
approaches do not easily carry over to~3D.

\begin{theorem}
  There exists an instance of 3D-MMN with $n$ terminals that requires
  a generating set of size $\Omega(n^2)$.
\end{theorem}

\begin{pf}
  We construct an instance that requires a generating set of size at
  least $n^2/4$. The main idea of the construction is to ensure that
  $n^2/4$ of the terminal pairs must use an edge segment unique to
  that specific pair.  The input consists of two sets $T$ and $T'$,
  each with $n/2$ terminals, with the following coordinates: for $0\le
  i < {n}/{2}$, terminal $t_i \in T$ is at $(i, {n}/{2} -i,
  {n}/{2} -i)$ and terminal $t'_i \in T'$ is at $({n}/{2}+i,
  n-i, n-i)$.  Figure~\ref{fig:genpairex} shows the instance for
  $n=6$.

  \begin{figure}
    \centering
    \includegraphics[scale=.4]{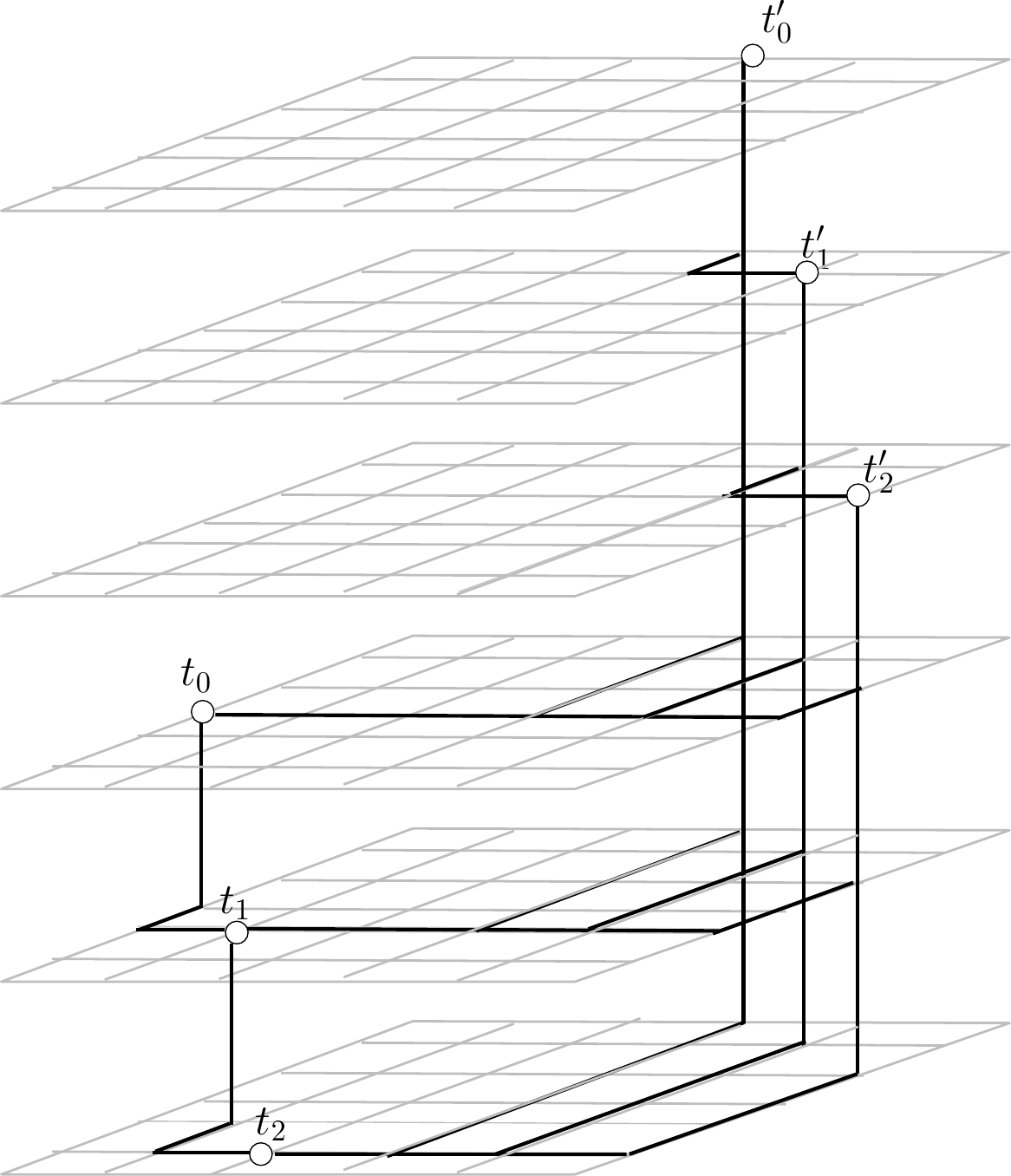}
    \caption{The constructed network for $n=6$.}
    \label{fig:genpairex}
  \end{figure}
  Consider any given generating set
  $Z \subset T \times T'$ such that there is a pair $(\tilde{t}, \tilde{t'})$, 
  $\tilde{t}\in T$ and $\tilde{t}'\in T'$ that is not in~$Z$.
  We now construct a specific network that contains M-paths between all 
  terminal pairs in $Z$ but no M-path between $(\tilde{t}, \tilde{t'})$.
  
  Consider any pair $(t,t')\in Z$ such that $t=(i,j,k) \in T$ and
  $t'=(i',j',k') \in T'$.  The
  M-path from $t$ to $t'$ has three segments: an $x$-aligned segment 
  from $(i,j,k)$ to $(i',j,k)$, a $y$-aligned segment 
  from $(i',j,k)$ to $(i',j',k)$, and a $z$-aligned segment 
  from $(i',j', k)$ to $(i',j',k')$. To ensure an M-path between
  each generating pair $t_i, t_j \in T$ (similarly between
  $t_i', t_j' \in T'$), we add M-paths between each pair of
  consecutive terminals in $T$ (similarly for $T'$) as follows:
  we connect $t_i, t_{i+1} \in T$ 
  by adding a $z$-aligned segment from $t_i=(i,j,k)$ to $(i,j,k-1)$, a
  $y$-aligned segment to $(i,j-1,k-1)$, and an $x$-aligned segment to
  $t_{i+1}=(i+1,j-1,k-1)$; see Fig.~\ref{fig:genpairex}.

  It is easy to verify that, in this construction, the M-path between
  terminals $t = (i,j,k) \in T$ and $t' = (i',j',k') \in T'$ must use
  the $y$-aligned segment between $(i',j,k)$ and
  $(i',j',k)$.  Since this segment is added only between terminal
  pairs that are present in the generating set~$Z$, there is no
  M-path between terminals $\tilde{t}\in T$ and $\tilde{t}'\in T'$
  which are not in $Z$. Thus, in order to obtain M-paths between all
  pairs of terminals in $T \cup T'$, we need at least all of the
  ${n^2}/{4}$ pairs in $T \times T'$.
\end{pf}

\subsection{Hanan Grid and Directional Subproblems}
\label{sec:hann-grid-direct}

First, we note that any instance of MMN has a solution that is
contained in the \emph{Hanan grid}, the grid induced by the terminals;
see Fig.~\ref{fig:examples}(a).  Gudmundsson \etal~\cite{gln-ammn-01}
showed this for~2D; their proof generalizes to higher dimensions.  In
what follows, we restrict ourselves to finding feasible solutions
that are contained in the Hanan grid.

Second, to simplify our proofs, we consider the \emph{directional}
subproblem of 3D-MMN which consists of connecting all terminal
pairs $(t,t')$ such that~$t$ \emph{dominates}~$t'$, that is, $x(t) \le
x(t')$, $y(t) \le y(t')$, $z(t) \le z(t')$, and $t \ne t'$.  We call such
terminal pairs \emph{relevant}.  

The idea behind our reduction to the directional subproblem is that
any instance of 3D-MMN can be decomposed into four subproblems of this
type.  One may think of the above-defined directional subproblem as
connecting the terminals which are oriented in a north-east (NE)
configuration in the $x$--$y$ plane (with increasing
$z$-coor\-di\-nates).  Analogous subproblems exist for the
directions~NW, SE, and~SW.  Note that any terminal pair belongs to one
of these four categories (if seen from the terminal with smaller
$z$-coordinate).

The decomposition extends to higher dimensions $d$, by fixing the
relationship between $(t,t')$ for one dimension (for example, $z$), and
enumerating over all possible relationships for the remaining $d-1$
dimensions. This decomposes $d$D-MMN into $2^{d-1}$ 
subproblems, which is a constant number of subproblems as we consider
$d$ to be a fixed constant.  

This means that any
$\rho$-approximation algorithm for the directional subproblem leads to
an $O(\rho)$-approxi\-mation algorithm for the general case.
Thus we can focus on designing algorithms for the directional subproblem.  

\begin{observation} 
  \label{obs:dir-sub} 
  Any instance of MMN can be decomposed into a constant number of
  directional subproblems.  Thus a $\rho$-approximation algorithm for
  the directional subproblem leads to an $O(\rho)$-approximation
  algorithm for MMN.
\end{observation}

\subsection{Relation to Steiner Problems}
\label{sec:relat-stein-type}

We next show that there is an approximation-preserving reduction
from directional 3D-MMN to the directed Steiner forest (DSF) problem,
which by Observation~\ref{obs:dir-sub}, carries over up to a constant
factor, to general 3D-MMN. 

Let $T$ be a set of $n$ points in $\mathbb{R}^3$.  Let $H$ be the
Hanan grid induced by $T$.  We consider $H$ as an undirected graph
where the length of each edge equals the Euclidean distance between
its endpoints.  We orient each edge in~$H$ so that, for any edge
$(p,p')$ in 
the resulting digraph~$H'$, the start node $p$ dominates the end node
$p'$.  We call $H'$ the \emph{oriented Hanan grid} of~$T$.  
Now let
$(t,t')$ be a relevant pair of points in $T$, that is, $t$
dominates~$t'$.  Any M-path in $H$ connecting $t$ to $t'$ corresponds
to a directed path in~$H'$ from~$t$ to~$t'$.  The converse also holds:
every directed path in~$H'$ corresponds to an M-path in~$H$.

Let $I$ be an instance of directional 3D-MMN and let $I'$ be an 
instance of DSF where the input graph is $H'$ and where every relevant
terminal pair of $I$ has to be connected.  Then, each feasible solution $N$
of $I$ contained in $H$ corresponds to a sub-graph~$N'$ of~$H'$ that
connects every relevant terminal pair, and is therefore a 
feasible solution to $I'$.  It is easy to see that $N'$ has the same
cost as $N$, 
as $N'$ uses the oriented version of each edge of $N$.
Conversely, every feasible solution~$N'$ for~$I'$ corresponds to a
subgraph $N$ of $H$ that M-connects every relevant terminal pair.
Therefore, $N$ is a feasible solution to $I'$ with the same cost as
$N'$.  This establishes an efficiently computable one-to-one
correspondence between feasible solutions to $I$ that are contained in
$H$ and feasible solutions to $I'$.  Since there is an optimum
solution to~$I$ contained in~$H$~\cite{gln-ammn-01}, this is an
approximation-preserving reduction from directional 3D-MMN to DSF.

By means of the above transformation of the Hanan grid into a digraph,
we also obtain an approximation-preserving reduction from 3D-RSA to
DST.  We use this %
later in
Section~\ref{sec:general-case} to develop an approximation algorithm
for 3D-MMN.  Let $I$ be an instance of 3D-RSA given by a set $T$ of
terminals with non-negative coordinates that are to be M-connected to
the origin $o$.  We construct an instance $I'$ of DSF as above where
$\{o\}\times T$ is the set of node pairs to be connected.
Note that any feasible solution to $I'$ is, without loss of
generality, a tree.  Hence, $I'$ is an instance of DST with root~$o$.
All in all, we have an approximation-preserving reduction from 3D-RSA
to DST. 

\section{The $k$-Plane Case}
\label{sec:k-planes}

In this section we consider 3D-MMN, under the assumption that the
set~$T$ of terminals is contained in the union of $k \ge 2$ planes
$E_1,\dots, E_k$ that are parallel to the $x$--$y$ plane.  Of course,
this assumption always holds for some $k \le n$.  We present a
$4(k-1)$-approximation algorithm, which outperforms our algorithm for
the general case in Section~\ref{sec:general-case} if $k \in
o(n^\eps)$.  

Let~\Nopt be some fixed minimum M-network for~$T$, let \Nopthor be the
set of all $x$-aligned and all $y$-aligned segments in~\Nopt, and let
\Noptver be the set of all $z$-aligned segments in~\Nopt.  Let~\opt
denote the weight of~\Nopt.  Clearly, \opt does not depend on the
specific choice of~\Nopt; the weights of~\Nopthor and~\Noptver,
however, may depend on~\Nopt. For $i \in \{1,\dots,k\}$, let $T_i =
T \cap E_i$ be the set of terminals in plane~$E_i$.  Further, let~\Txy
be the projection of~$T$ onto the $x$--$y$ plane. 

Our algorithm consists of two phases.  Phase~I computes a set~\Nhor of
horizontal (that is, $x$- and $y$-aligned) line segments, phase~II
computes a set~\Nver of vertical (that is, $z$-aligned) line segments.
Finally, the algorithm returns the set $N = \Nhor \cup \Nver$. 

Phase~I is simple; we compute a 2-approximate M-network~\Nxy for~\Txy
(using the algorithm of Guo \etal~\cite{gsz-yaa2a-08}) and
project~\Nxy %
onto each of the planes $E_1,\dots, E_k$.  Let~\Nhor be the union of
these projections.  Note that \Nhor M-connects any pair of terminals
that lie in the same plane.
\begin{observation}
  \label{obs:Nhor}
  $\| \Nhor \| \le 2k\|\Nopthor\|$.
\end{observation}
\begin{pf}
  The projection of~\Nopthor to the $x$--$y$ plane is an M-network
  for~\Txy.  Hence, $\|\Nxy\| \le 2 \Nopthor$.  Adding up over the $k$
  planes yields the claim.
\end{pf}
In Phase~II, we construct a {\em pillar network} by computing a set~\Nver of
vertical line segments, so-called \emph{pillars}, of total cost at
most $4(k-1)\|\Noptver\|$.  This yields an overall approximation
factor of~$4(k-1)$ since $\|\Nhor \cup \Nver\| \le 2k\|\Nopthor\| +
4(k-1)\|\Noptver\| \le 4(k-1) (\|\Nopthor\| + \|\Noptver\|) \le 4(k-1)
\opt$.

Below we describe Phase~II of our algorithm 
for the directional subproblem 
that runs in direction north-east (NE) in the $x$--$y$
plane (with increasing $z$-coordinates). For this directional subproblem, we
construct a pillar
network~$\Nver_\dir$ of weight at most $(k-1)\|\Noptver\|$ that, together
with~\Nhor, M-connects all relevant pairs.   
We solve the analogous
subproblems for the directions~NW, SE, and~SW in the same fashion.
Then \Nver is the union of the four partial solutions and has weight
at most $4(k-1)\|\Noptver\|$, as desired.

Our directional subproblem is closely linked to the
\emph{(directional) bichromatic rectangle piercing problem} (\BRP),
which is defined as follows.  Let~\red and~\blue be sets of red and
blue points in $\R^2$, respectively, and let ${\cal R}(\red,\blue)$
denote the set of axis-aligned rectangles each of which is spanned by
a red point in its SW-corner and a blue point in its NE-corner.  Then
the aim of \BRP is to find a minimum-cardinality set~$P \subset \R^2$
such that every rectangle in~${\cal R}(\red,\blue)$ is \emph{pierced},
that is, contains at least one point in~$P$.  The points in~$P$ are
called \emph{piercing points}. %

The problem dual to BRP is the \emph{(directional) bichromatic
  independent set of rectangles problem} (BIS) where the goal is to
find the maximum number of pairwise disjoint rectangles in ${\cal
  R}(\red,\blue)$, given the sets~\red and~\blue.

Recently, Soto and Telha \cite{st-2dorg-11} proved a beautiful
min--max theorem saying that, for ${\cal R}(\red,\blue)$, the
minimum number of piercing points always \emph{equals} the maximum
number of independent rectangles.  This enabled them to give efficient
exact algorithms for \BRP and BIS running in $\tilde{O}(n^{2.5})$
worst-case time or $\tilde{O}(n^{\gamma})$ expected time, where the
$\tilde{O}$-notation ignores polylogarithmic factors, $\gamma <
2.4$ is the exponent for fast matrix multiplication, and
$n=|\red|+|\blue|$ is the input size.

The details of Phase~II appear, for $k=2$ planes, in Section~\ref{sec:pillar2},
and, for $k>2$ planes, in Section~\ref{sec:pillark}.
Algorithm~\ref{alg:kplanes} summarizes of our $k$-planes algorithm.

\begin{algorithm}[htpb]
\caption{$k$-Planes Algorithm}
\label{alg:kplanes}
\mbox{Input: Set $T$ of terminals contained in the union of planes $E_1,
\ldots, E_k$, all parallel to the $x$--$y$ plane.}\\[-2.5ex]
\begin{algorithmic}[1]
\STATE Let $T_{xy}$ be the projection of $T$ onto the $x$--$y$ plane 
\STATE Phase I: 
\\\quad Compute $\Nxy$, a 2-approximate M-network for $T_{xy}$ using the algorithm of
Guo \etal~\cite{gsz-yaa2a-08}.
\\\quad Let $\Nhor$ be the union of the projections of $N_{xy}$ onto each
of the planes $E_1, \ldots, E_k$.
\STATE Phase II:  
\\\quad If $k=2$, construct a pillar network $\Nver$ by
Algorithm~\ref{alg:2pillars}; see Section~\ref{sec:pillar2}.
\\\quad Otherwise, construct a pillar network $\Nver$ by
Algorithm~\ref{alg:kpillars}; see Section~\ref{sec:pillark}.
\end{algorithmic}
Output: $\Nhor \cup \Nver$.
\end{algorithm}

\subsection{Pillar Network for Two Planes}
\label{sec:pillar2}

Our phase-II algorithm for two planes is very simple.  We sketch it
first in order to provide some intuition for the $k$-planes
case.  Let the terminals in~$T_1$ be red and those in~$T_2$  be blue.
Ignore the $z$-coordinates of the terminals.  Then the relevant
red--blue point pairs span exactly the rectangles in ${\cal
  R}(T_1,T_2)$, which we call relevant, too.

\begin{algorithm}[htpb]\caption{Pillar network of the directional subproblem for
$k=2$ planes}
Input: Sets $T_1 \subset E_1$ and $T_2 \subset E_2$ of terminals.
\label{alg:2pillars}
\begin{algorithmic}[1]
\STATE Color $T_1$ red and $T_2$ blue.
\STATE Ignoring $z$-coordinates of terminals, let ${\cal R}(T_1,T_2)$ be
the set of rectangles spanned by relevant red--blue pairs.  
\STATE Compute a minimum piercing~$\hat{P}$ of ${\cal R}(T_1,T_2)$ such that
for each relevant red--blue pair $(r,b) \in T_1 \times T_2$ the piercing point
for $(r,b)$ lies on an $r$--$b$ M-path in $N_{xy}$, as described in Lemma
\ref{lem:paths}.
\STATE Erect pillars from $E_1$ to $E_2$ at each piercing point $\hat{p} \in
\hat{P}$; let $\Nver_\dir$ be the resulting set of pillars. 
\end{algorithmic}
Output: $\Nver_\dir$.
\end{algorithm}

Our algorithm (Algorithm~\ref{alg:2pillars}) consists of two steps.
First, we compute a minimum 
piercing~$P$ of ${\cal R}(T_1,T_2)$ using the algorithm of Soto and
Telha~\cite{st-2dorg-11}.  Second, we move each piercing point~$p \in
P$ to a new position $\hat{p}$---a nearby junction of~\Nxy---and
erect, at~$\hat{p}$, a pillar connecting the two planes.
Let~$\hat{P}$ be the set of piercing points after the move, and let
$\Nver_\dir$ be the corresponding set of pillars.

\begin{lemma}
  \label{lem:4approx}
  It holds that $\|\Nver_\dir\| \le \|\Noptver\|$.
\end{lemma}

\begin{pf}
  It is easy to see that $|\hat{P}|=|P|$.  Integrating over the
  distance~$d$ of the 
  two planes yields $\|\Nver_\dir\| = |\hat{P}| \cdot d = |P| \cdot d
  \le \|\Noptver\|$.  The last inequality is due to the fact that $P$
  is a \emph{minimum} piercing of ${\cal R}(T_1,T_2)$ and that the pillars
  in~\Noptver pierce ${\cal R}(T_1,T_2)$---otherwise~\Nopt would not
  be feasible.
\end{pf}

Now we turn to feasibility.  We first detail how we move each piercing
point~$p$ to its new position~$\hat{p}$.  For the sake of brevity, we
identify terminals with their projections to the $x$--$y$ plane.  Our
description assumes that we have at our disposal some network~$M$
(such as~\Nxy) connecting the relevant pairs in~\Txy. 

\begin{figure}
  \centering
  \includegraphics{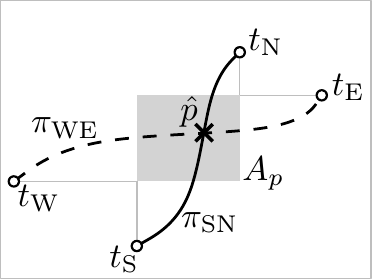}
  \caption{Paths~\MSN and~\MWE meet in a point~$\hat{p}$ in~$A_p$.}
  \label{fig:piercing}
\end{figure}

For a piercing point $p \in P$, let $A_p$ be the intersection
of the relevant rectangles pierced by~$p$; see
Fig.~\ref{fig:piercing}.  Clearly, $p \in A_p$.  Note that the bottom
and left  
sides of~$A_p$ are determined by terminals~\tW and~\tS to the west and
south of~$A_p$, respectively.  Symmetrically, the top and right sides
of~$A_p$ are determined by terminals~\tE and~\tN to the east and north
of~$A_p$, respectively.  Terminals~\tW and~\tS may coincide,
and so may~\tE and~\tN.  It is easy to see that the network~$M$
contains an M-path~\MSN 
connecting~\tS and~\tN and an M-path~\MWE connecting~\tW and~\tE.  The
path~\MSN goes through the bottom and top sides of~$A_p$ and~\MWE goes
through the left and right sides.  Hence, the two paths intersect in a
point~$\hat{p} \in A_p$.  This is where we move the original piercing
point~$p$.  

Since~$\hat{p} \in A_p$, the point~$\hat{p}$ pierces
the same relevant rectangles as~$p$, and the set
$\hat{P} = \{ \hat{p} \mid p \in P \}$ is a (minimum) piercing for the
set of relevant rectangles. 

\begin{lemma}
  \label{lem:paths}
  Let ${\cal R}(\red,\blue)$ be an instance of \BRP and let $M$ be a
  network that M-connects every relevant red--blue point pair.  Then
  we can efficiently compute a minimum piercing of ${\cal
    R}(\red,\blue)$ such that~$M$ contains, for every relevant
  red--blue point pair~$(r,b)$ in~$\red \times \blue$, an $r$--$b$
  M-path that contains a piercing point.
\end{lemma}

\begin{pf}
  We use the algorithm of Soto and Telha~\cite{st-2dorg-11} to compute
  a minimum piercing~$P$ of~${\cal R}(\red,\blue)$.  Then, as we have
  seen above, $\hat{P}$ is a minimum piercing of~${\cal
    R}(\red,\blue)$, too.  Now let $(r,b)$ be a relevant red--blue
  pair in~$\red\times\blue$, and let $p \in P$ be a point that
  pierces~$R(r,b)$.  Clearly, $\hat{p}$ pierces~$R(r,b)$, too.
  As we have observed before, both~$p$ and~$\hat{p}$ lie in~$A_p$.

  Since $(r,b)$ is a relevant pair, $r$ lies to the~\SW of~$A_p$
  and~$b$ to the~\NE; see Fig.~\ref{sfg:rin}. 
  We prove that $M$ contains an $r$--$\hat{p}$ M-path; a symmetric
  argument proves that~$M$ also contains a $\hat{p}$--$b$ M-path.
  Concatenating these two M-paths yields the
  desired $r$--$b$ M-path since~$r$ lies to the~\SW of~$\hat{p}$
  and~$\hat{p}$ lies to the~\SW of~$b$.  Recall that~$\hat{p}$ lies
  on the intersection of the \tW--\tE M-path \MWE and the \tS-\tN
  M-path \MSN, where \tW, \tE, \tS, \tN are the terminals that
  determine the extensions of~$A_p$; see Fig.~\ref{fig:piercing}.  To
  show that~$M$ M-connects~$r$ and~$\hat{p}$, we consider two
  cases.

  \smallskip
  \emph{Case I:} $r \in R(\tW,\tS)$;
  see Fig.~\ref{sfg:rin}.
  According to our assumption, $M$ contains \emph{some} $r$--$b$
  M-path~$\pi$. Then $\pi$ must intersect~\MWE
  or~\MSN at some point~$x$ to the~\SW of~$\hat{p}$.  Thus, we can go,
  in a monotone fashion, along~$\pi$ from~$r$ to~$x$ and then along~\MWE
  or~\MSN from~$x$ to~$\hat{p}$.  This is the desired $r$--$\hat{p}$
  M-path.

\begin{figure}
  \null\hfill
  \subfloat[Case~I: we can go from~$r$ via~$x$ to~$\hat{p}$.%
  \label{sfg:rin}]%
  {\qquad\includegraphics{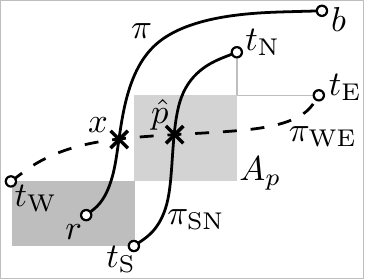}\qquad}
  \hfill
  \subfloat[Case~II: we can go from~$r$ via~\tW or~\tS to~$\hat{p}$.%
  \label{sfg:rout}]%
  {\qquad\includegraphics{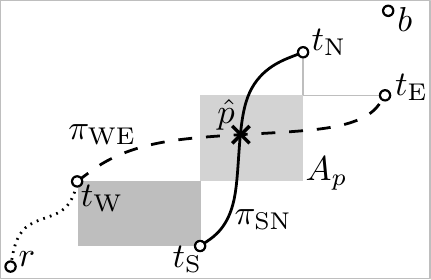}\qquad}
  \hfill\null

  \caption{Sketches for the proof of Lemma~\ref{lem:paths}.}
  \label{fig:paths}
\end{figure}

  \smallskip
  \emph{Case II:} $r$ lies to the~\SW of~\tW or~\tS; 
  see Fig.~\ref{sfg:rout}.
  In this case $M$ contains M-paths from~$r$ to~\tW and to~\tS.
  If~$r$ lies to the~\SW of~\tW, we can go, again in a monotone
  fashion, from~$r$ to~\tW and then along~\MWE from~\tW to~$\hat{p}$.
  Otherwise, if~$r$ lies to the~\SW of~\tS, we can go from~$r$
  to~\tS and then on~\MSN from~\tS to~$\hat{p}$.

  Since these are the only two possibilities, this concludes the proof.
\end{pf}

Lemmas~\ref{lem:4approx} and~\ref{lem:paths} (with $\red=T_1$,
$\blue=T_2$, and $M=\Nxy$) yield the following. %
\begin{theorem}
  \label{thm:select-pill-locat}
  We can efficiently compute a 4-approximation for the 2-plane~case.
\end{theorem}

\subsection{Pillar Network for $k$ Planes}\label{sec:pillark}

Now we show how our phase-II algorithm generalizes to $k$ planes. 
As in the 2-planes case, we restrict ourselves to the 
directional subproblem and construct a pillar network $\Nver_\dir$ of weight at
most $(k-1) \|\Noptver\|$.  As we have argued at the
beginning of Section~\ref{sec:k-planes}, this suffices to prove
Theorem~\ref{thm:k-planes}.  

\begin{theorem}
  \label{thm:k-planes}
  There exists a $4(k-1)$-approximation algorithm for 3D-MMN
  where the terminals lie in the union of $k \ge 2$ planes parallel to
  the $x$--$y$ plane.
\end{theorem}

Our pillar-placement algorithm (Algorithm~\ref{alg:kpillars}) is as
follows.  Let 
$i \in \{1,\dots,k-1\}$.  We construct an instance~${\cal I}_i$ of \BRP
where we two-color \Txy such that each point
corresponding to a terminal of some plane~$E_j$ with $j\leq i$ is
colored red and each point corresponding to a terminal of some
plane~$E_{j'}$ with $j'\geq i+1$ is colored blue.  For~${\cal I}_i$, we
compute a minimum piercing~$\hat{P}_i$ according to
Lemma~\ref{lem:paths} with $M=\Nxy$.  In other words, for any relevant
pair $(t_j,t_{j'})$, there is some M-path in $N_{xy}$ that contains a
piercing point of~$\hat{P}_i$.  We choose $i^\star \in \{1,\dots,k-1\}$
such that~$\hat{P}_{i^\star}$ has minimum cardinality.  This is
crucial for our analysis.
At the piercing points of~$\hat{P}_{i^\star}$, we erect pillars
spanning all planes $E_1,\ldots,E_k$.  Let~$\hat{N}_{i^\star}$ be the
set of these pillars.  We now show that
$\hat{N}_{i^\star}$, along with \Nhor, creates a feasible network for
any relevant terminal pair $(t_j, t_j')$ such that $j\le i^{\star}$
and $j'\ge i^{\star}+1$. 

\begin{algorithm}[htpb]
\caption{Pillar network for the directional subproblem for $k>2$ planes}
\label{alg:kpillars}
Input: Sets $T_s \subset E_s, \dots, T_t \subset E_t$ of terminals
with $s\le t$ (initially $s=1$ and $t=k$).   
\begin{algorithmic}[1]
\STATE Let $T'$ be the projection of $T_s \cup \dots \cup T_t$ onto
the $x$--$y$ plane. 
\FOR{\textbf{each} $i \in \{s,\dots,t\}$}
\STATE Let~${\cal I}_i$ be an instance of \BRP where each point in
$T'$, corresponding to a terminal in~$T_j$ with $j\leq i$, is 
colored red and each point in $T'$, corresponding to a
terminal in~$T_{j'}$ with $j'\geq i+1$, is colored blue. 
\STATE Compute a minimum piercing~$\hat{P}_i$ according to Lemma~\ref{lem:paths} with $M=\Nxy$.
\ENDFOR
\STATE Choose $i^\star \in \{s,\dots,t\}$ such that~$\hat{P}_{i^\star}$ has minimum cardinality. 
\STATE Let $\hat{N}_{i^\star}$ be the set of pillars erected at each
piercing point of~$\hat{P}_{i^\star}$, spanning planes $E_s, \dots, E_t$. 
\STATE Let $\hat{N}_{\le i^\star}$ be the output of this algorithm
applied recursively to $T_s, \dots, T_{i^{\star}}$.
\STATE Let $\hat{N}_{> i^\star}$ be the output of this algorithm
applied recursively to $T_{i^{\star}+1}, \dots, T_t$.
\end{algorithmic}
Output: $\hat{N}_{i^\star} \cup \hat{N}_{\le i^\star} \cup \hat{N}_{> i^\star}$
\end{algorithm}

\begin{lemma}
  \label{lem:feasibility}
  The network $\Nhor \cup \hat{N}_{i^\star}$ M-connects any relevant
  terminal pair in $T_j \times T_{j'}$ with $j \le i^\star$ and 
  $j' \ge i^\star+1$.
\end{lemma}

\begin{pf}
  Consider a pair $(t_j,t_{j'})$ in $T_j \times T_{j'}$ as in the
  statement.  We construct an M-path from~$t_j$ to~$t_{j'}$ as
  follows.  We know that there exists an M-path~$\pi$ that connects the
  projections of~$t_j$ and~$t_{j'}$ in~\Nxy and contains a piercing
  point~$p$ of~$\hat{P}_{i^\star}$.  Therefore, we can start at~$t_j$
  and follow the 
  projection of~$\pi$ onto plane~$E_j$ until we arrive at~$p$.  Then we
  use the corresponding pillar in~$\hat{N}_{i^\star}$ to reach the
  plane~$E_{j'}$, where we follow the projection of~$\pi$ (onto that
  plane) until we reach~$t_{j'}$.  
\end{pf}

In order to also M-connect relevant terminal pairs in $T_j \times T_{j'}$, where
either ($j \le i^\star$ and $j' \le i^\star$) or ($j \ge i^\star+1$
and $j' \ge i^\star+1$), we simply apply
the pillar-placement algorithm recursively to the sets
$T_1,\dots,T_{i^\star}$ and $T_{i^\star+1},\dots,T_k$. 
This yields the desired pillar network~$\Nver_\dir$.  By
Lemma~\ref{lem:feasibility}, $\Nver_\dir \cup \Nhor$ is feasible.
Next, we bound $\|\hat{N}_{i^\star}\|$.

\begin{lemma}
  \label{lem:piercing-split}
  Let~$M$ be an
  arbitrary directional Manhattan network for~$T$, and let~\Mver be 
  the set of vertical segments in~$M$.  Then the pillar network
  $\hat{N}_{i^\star}$ has weight at most $\|\Mver\|$.  
\end{lemma}

\begin{pf}
  Without loss of generality, we assume that $M$ is a subnetwork of 
  the Hanan grid~\cite{gln-ammn-01}.  We may also assume that any
  segment of~\Mver spans only 
  consecutive planes. %
  For $1\leq i\leq j\leq k$, let $M_{i,j}$
  denote the subnetwork of~\Mver lying between planes $E_i$ and $
  E_j$.  Let $d_{i,j}$ be the vertical distance between planes~$E_i$
  and~$E_j$.

  We start with the observation that, for any $j=1,\ldots,k-1$, the
  network~$M_{j,j+1}$ is a set of pillars that forms a valid piercing
  of the piercing instance~${\cal I}_j$ (defined right after
  Theorem~\ref{thm:k-planes}).  Hence, $|M_{j,j+1}| \ge |\hat{P}_j| 
  \ge |\hat{P}_{i^\star}|$, which implies the claim of the lemma
  as follows:
  \begin{displaymath}
    \|\Mver\|=\sum_{j=1}^{k-1}\|M_{j,j+1}\|=\sum_{j=1}^{k-1}|M_{j,j+1}|
    \cdot d_{j,j+1} \ge \sum_{j=1}^{k-1}|P_{i^\star}| \cdot d_{j,j+1} =
    |P_{i^\star}| \cdot d_{1,k} = 
    \|P_{i^\star}\|.   
  \end{displaymath}
\end{pf}

It is crucial for our construction that the pillars constructed
recursively span either 
$E_1,\dots,E_{i^\star}$ or $E_{i^\star+1},\dots,E_k$, but not all
planes.
For $1\leq j\leq j'\leq k$, let $\cost_z(j,j')$ denote the weight of
the vertical part of the network produced by the above
pillar-placement algorithm, when applied to planes $E_j,\dots,E_{j'}$
recursively.  For technical reasons we set ${\cost_z(j,j)=0}$.  Now assume that $j<j'$ and that the algorithm makes the partition at
plane $E_{i'}$ with $j\leq i'< j'$ when planes $E_j,\dots,E_{j'}$ are
processed.  By means of Lemma~\ref{lem:piercing-split}, we derive the
recursion 
\begin{equation}
  \label{eqn:weight}
  \cost_z(j,j')\leq \|M_{j,j'}\|+\cost_z(j,i')+\cost_z(i'+1,j')\, ,
\end{equation}
which holds for any M-network~$M$ for~$T$.  We now claim that
\begin{displaymath}
  \cost_z(j,j')\leq (j'-j)\|M_{j,j'}\|.
\end{displaymath}
Our proof is by induction on the number of planes processed by the
algorithm.  By the inductive hypothesis, we have that
$\cost_z(j,i')\leq(i'-j)\|M_{j,i'}\|$ and
$\cost_z(i'+1,j')\leq(j'-i'-1)\|M_{i'+1,j'}\|$.  
We plug these expressions into the recursion~\ref{eqn:weight}.
Since
$\|M_{j,i'}\|+\|M_{i'+1,j'}\|\leq \|M_{j,j'}\|$ and 
$\cost_z(l,l)=0$ for any $l\in\{1,\dots,k\}$, the claim follows.  

We conclude that the
weight of the solution produced by the algorithm, when applied to all
planes $E_1,\dots,E_k$, is bounded by $\cost_z(1,k)\leq
(k-1)\|M_{1,k}\|=(k-1)\|\Mver\|$.
This completes the proof of Theorem~\ref{thm:k-planes}.

\section{The General Case}
\label{sec:general-case}

In this section, we present an
approximation algorithm, which we call the \emph{grid 
  algorithm}, for the general 3D-MMN problem.  Our main result is the following.
\begin{theorem}
  \label{thm:3dmain}
  For any $\eps>0$, there exists an $O(n^\eps)$-approximation
  algorithm for 3D-MMN that, given a set of $n$ terminals, runs in
  $n^{O(1/\eps)}$ time.
\end{theorem}
This result is better than the one in the previous section if the
given set of terminals is distributed over $\omega(n^\eps)$ horizontal
planes.  Moreover, the approach in this section extends to higher
dimensions; see appendix.

For technical reasons, we assume that the terminals are in general
position, that is, any two terminals differ in all three coordinates.
By Observation~\ref{obs:dir-sub} it suffices to describe and analyze
the algorithm for the directional subproblem. 

\subsection{The 3D Grid Algorithm} 
\label{sec:gridalg}

\begin{figure}[h]
  \subfloat[Example of a grid with two cuboids from
  different slabs.\label{sfg:gridedges}]%
  {\parbox{.3\textwidth}{\centering\includegraphics[width=\linewidth]{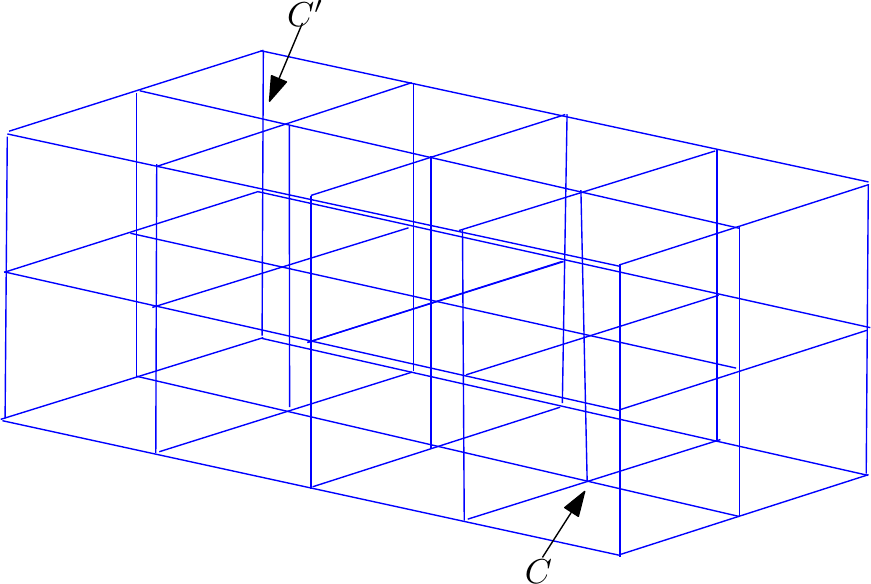}}}
  \hfill
  \subfloat[Examples of $x$-, $y$-, and $z$-aligned slabs.\label{sfg:slabs}]%
  {\parbox{.35\textwidth}{\centering\includegraphics[width=\linewidth]{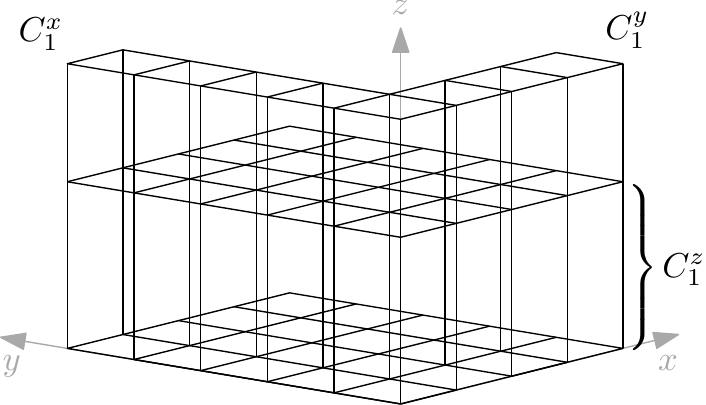}}}
  \hfill
  \subfloat[M-paths using patching and grid edges.\label{sfg:patchingpath}]
  {\parbox{.25\textwidth}{\centering\includegraphics[width=\linewidth]{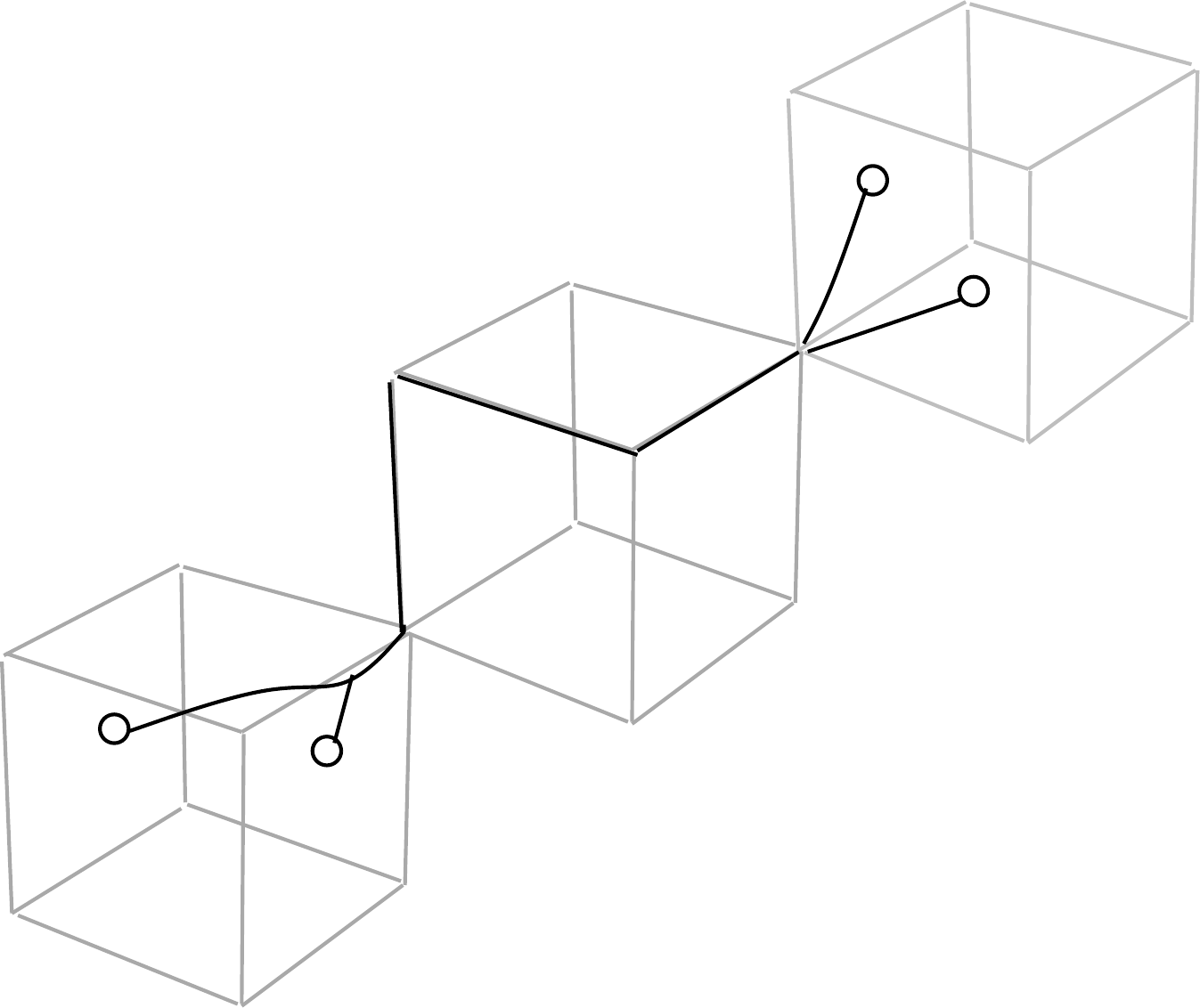}}}
  \caption{Illustrations for the grid algorithm: cuboids, slabs, and patches.}
  \label{fig:alg3d}  
\end{figure}

We begin the description with a high-level summary.  To solve the
directional subproblem, we place a 3D grid that partitions the
instance into a constant number of cuboids; see
Fig.~\ref{sfg:gridedges}. Cuboids that differ in only two 
coordinates form \emph{slabs}. We connect terminals from different
slabs by M-connecting each terminal to the corners of its cuboid and
by using the edges of the grid to connect the corners.  We connect
terminals from the same slab by recursively applying our algorithm to
the slabs.

\medskip
\noindent
{\em Step 1: Partitioning into cuboids and slabs.} Consider the bounding
cuboid~$C$ of~$T$ and set $c=3^{1/\eps}$.  Partition~$C$ 
by $3(c-1)$ separating planes into $c\times c\times c$ axis-aligned
subcuboids $C_{ijk}$ with $i,j,k \in \{1,\ldots,c\}$.  The indices are
such that larger indices mean larger coordinates.  Place the separating
planes such that the number of terminals between two consecutive planes is at
most $n/c$.  This can be accomplished by executing a simple plane-sweep 
for each direction $x, y, z,$ and by placing separating planes after
every~$n/c$ terminals.  Here we exploit our general-position assumption.
The edges of the resulting subcuboids---except the edges on the boundary
of~$C$, which we do not need---induce a three-dimensional 
grid~$\mathcal{G}$ of axis-aligned line segments.  We insert~$\mathcal{G}$
into the solution.

For each $i \in \{1,\dots,c\}$, define the $x$-aligned \emph{slab},
$C_i^x$, to be the union of all cuboids $C_{ijk}$ with $j,k \in
\{1,\dots,c\}$.  Define $y$-aligned and $z$-aligned
slabs $C_j^y$, $C_k^z$ analogously; see Fig.~\ref{sfg:slabs}.

\medskip
\noindent
{\em Step 2: Add M-paths between different slabs.} 
Consider two cuboids $C_{ijk}$ and $C_{i'j'k'}$ with $i<i'$, $j<j'$,
and $k<k'$.  Any terminal pair $(t,t') \in C_{ijk} \times C_{i'j'k'}$
can be M-connected using the edges of~$\mathcal G$ as long 
as~$t$ and~$t'$ are connected to the appropriate corners of their
cuboids; see Fig.~\ref{sfg:patchingpath}.  To this end, we use the
following \emph{patching} procedure. 

Call a cuboid $C_{ijk}$ \emph{relevant} if there is a non-empty cuboid
$C_{i'j'k'}$ with $i<i'$, $j<j'$, and $k<k'$.  For each relevant
cuboid~$C_{ijk}$, let~$\hat{p}_{ijk}$ denote a corner that is dominated by all
terminals inside~$C_{ijk}$.
We define \emph{up-patching}~$C_{ijk}$ to mean M-connecting every terminal
in~$C_{ijk}$ to~$\hat{p}_{ijk}$.  We up-patch~$C_{ijk}$ by solving 
(approximately) an instance of 3D-RSA with the terminals in~$C_{ijk}$
as points and $\hat{p}_{ijk}$ as origin.  We define \emph{down-patching}
analogously; cuboid $C_{ijk}$ is relevant if there is a non-empty
cuboid $C_{i'j'k'}$ with $i>i'$, $j>j'$, $k>k'$; we let $\check{p}_{ijk}$
be the corner that dominates all terminals in~$C_{ijk}$.

We complete this step by inserting the up-patches and the down-patches
of all relevant cuboids into the solution. 

\medskip
\noindent
{\em Step 3: Add M-paths within slabs.}  
To M-connect relevant terminal pairs that lie in the same slab, we
apply the grid algorithm (steps 1--3) recursively to each slab $C_i^x$,
$C_j^y$, and $C_k^z$ with $i,j,k \in \{1,\dots,c\}$.

\subsection{Analysis}

We first show that the output of the algorithm presented in
Section~\ref{sec:gridalg} is feasible, then we 
establish its approximation ratio of $O(n^\eps)$ and its running time
of $n^{O(1/\eps)}$ for any $\eps > 0$. 
In this section,
\opt denotes the weight of a minimum M-network (\emph{not} the cost of
an optimal solution to the directional subproblem).

\begin{lemma}[Feasibility]
  \label{lem:mpaths}
  The grid algorithm M-connects all relevant terminal pairs.
\end{lemma}

\begin{pf} 
  Let $(t,t')$ be a relevant terminal pair.  
  First, suppose that $t$ and $t'$ lie in cuboids of different slabs.
  Thus, there are $i<i',j<j',k<k'$ such that $t\in C_{ijk}$ and $t'\in
  C_{i'j'k'}$.  Furthermore, $C_{ijk}$ and $C_{i'j'k'}$ are
  relevant for up- and down-patching, respectively.
  When up-patching, we solve an instance of RSA
  connecting all terminals in~$C_{ijk}$ to~$\hat{p}_{ijk}$.  Similarly,
  down-patching M-connects~$t'$ to~$\check{p}_{i'j'k'}$.
  The claim follows as $\cal G$ M-connects~$\hat{p}_{ijk}$
  and~$\check{p}_{i'j'k'}$.

  Now, suppose that~$t$ and~$t'$ lie in the same slab.  As the
  algorithm is applied recursively to each slab, there will be a recursion
  step where~$t$ and~$t'$ lie in cuboids in different slabs.  Here, we
  need our general-position assumption.  Applying the argument above
  to that particular recursive step completes the proof.
\end{pf}

\paragraph{Approximation ratio.}

Next, we turn to the performance of our algorithm.  Let $r(n)$ be its
approximation ratio, where $n$ is the number of terminals in $T$.  The
total weight of the output is the sum of~$\|{\cal G}\|$, the cost of
patching, and the cost for the recursive treatment of the slabs.  We
analyze each of the three costs separately.

The grid~$\mathcal G$ consists of all edges induced by the $c^3$
subcuboids except the edges on the boundary of~$C$.  Let $\ell$ denote
the length of the longest side of~$C$.  The weight of~$\cal G$ is at
most $3(c-1)^2 \ell$, which is bounded by $3c^2\opt$ as $\ell \le \opt$.

Let $\rpatch(n)$ denote the cost of patching all relevant
cuboids in step~2.  Lemma~\ref{lem:dstpatch} (given below) proves that
$\rpatch(n) = O(n^{\eps})\opt$.

Now consider the recursive application of the algorithm to all slabs.
Recall that $\Nopt$ is a fixed minimum M-network for~$T$.  For $i \in
{1,\ldots,c}$, let $\opt_i^x$ be the optimum cost for M-connecting \emph{all}
(not only relevant) terminal pairs in slab~$C_i^x$. Define~$\opt_i^y$
and~$\opt_i^z$ analogously. 

Slightly abusing of notation, we write $\Nopt\cap C_i^x$ for the set
$\{ s \cap C_i^x \mid s \in \Nopt\}$
of line segments of~\Nopt intersected with
slab~$C_i^x$.  Observe that $\Nopt\cap C_i^x$ forms a feasible
solution for~$C_i^x$.  Thus, $\opt_i^x\leq \|\Nopt\cap C_i^x\|$.  By
construction, any slab contains at most $n/c$ terminals.  Hence, the
total cost of the solutions for slabs $C_1^x,\dots,C_c^x$ is at most
\begin{displaymath}
  \sum_{i=1}^cr\left(\frac{n}{c}\right)\opt_i^x\leq
  r\left(\frac{n}{c}\right)\sum_{i=1}^c\|\Nopt\cap C_i^x\|\leq
  r\left(\frac{n}{c}\right)\opt\,.
\end{displaymath}
Clearly, the solutions for the $y$- and $z$-slabs have the same bound.

Summing up all three types of costs, we obtain the recursive equation
\begin{displaymath}
  r(n)\opt\le 3c^2\opt+\rpatch(n)\opt+3r\left(\frac{n}{c}\right)\opt\,.
\end{displaymath}
Hence, $r(n)=O(n^{\max\{\eps,\log_c3\}})$.  Plugging in $c =
3^{1/\eps}$ yields $r(n)=O(n^\eps)$, which proves the approximation
ratio claimed in Theorem~\ref{thm:3dmain}.

\begin{lemma}
  \label{lem:dstpatch} 
  Patching all relevant cuboids costs $\rpatch(n) \in O(n^{\eps})\opt$.
\end{lemma}

\begin{pf}
  First note that it suffices to consider up-patching; the
  down-patching case can be argued analogously. 

  Lemma~\ref{lem:patching-cost} shows the existence of a near-optimal
  M-network that up-patches all relevant cuboids.
  Lemma~\ref{lem:patching-algorithm} shows that by reducing the
  patching problem to 3D-RSA, we can find such a network of cost
  $O(\rho)\opt$, where $\rho$ is the approximation factor of 3D-RSA.

  We argued in Section~\ref{sec:relat-stein-type} that there exists an
  approximation-preserving reduction from 3D-RSA to DST.  DST, in
  turn, admits an $O(n^\eps)$-approximation for any $\eps>0$
  \cite{cccdggl-aadsp-98}.  Hence, the cost of up-patching is indeed
  bounded by $O(n^\eps)\opt$.
\end{pf}

We now turn to the two lemmas that we just used in the proof of
Lemma~\ref{lem:dstpatch}.  For our analysis, we need the network~$N'$
that is the union of~$\mathcal G$ with~\Nopt and the projections
of~\Nopt onto every separating plane of~$\mathcal G$.  Since there are
$3(c-1)$ separating planes and, as we have seen above, $\|{\cal G}\|
\le 3c^2\opt$, it holds that $\|N'\| \le 3(c^2+c)\opt = O(\opt)$.

\begin{lemma}
  \label{lem:patching-cost}
  There exists an M-network of total cost at most $3(c^2+c)\opt$
  that up-patches all relevant cuboids.
\end{lemma}

\begin{pf}
  We claim that $N'$ up-patches all relevant cuboids.  To this end,
  let $t\in C_{ijk}$ and let $t'\in C_{i'j'k'}$ with $i<i',j<j',k<k'$.
  Follow the M-path connecting $t$ and $t'$, starting from $t$.  This
  path must leave $C_{ijk}$ at a certain point $\bar p$, which lies on
  some face $F$ of $C_{ijk}$.  Face $F$, in turn, lies on some
  separating plane $S$ of the grid $\mathcal G$.  From now on follow the
  projection of the M-path from $\bar p$ to $t'$ on plane $S$.  This
  projected path must leave the face $F$, since $t'$ lies in
  $C_{i'j'k'}$ with $i<i',j<j',k<k'$, and the projection of $t'$ onto
  $S$ must therefore lie outside of~$F$.  Moreover, the point $\bar p'$
  where this path leaves~$F$ must lie on an edge of~$C_{ijk}$ incident
  to~$p_{ijk}$.  Hence, we obtain a $t$--$p_{ijk}$ M-path by going
  from~$t$ to~$\bar p$, from~$\bar p$ to~$\bar p'$ and then from~$\bar
  p'$ to~$p_{ijk}$.
\end{pf}

\begin{lemma}
  \label{lem:patching-algorithm}
  Given a number $\rho \ge 1$ and an efficient $\rho$-approximation of
  3D-RSA, we can efficiently up-patch all relevant cuboids at cost no
  more than $12(c^2+c) \rho\opt$.
\end{lemma}

\begin{pf}
  In Lemma~\ref{lem:patching-cost}, we showed the existence of a
  network $N'$ that up-patches all relevant cuboids at low cost.  Now
  consider an arbitrary relevant cuboid $C_{ijk}$.  Clearly $N'\cap
  C_{ijk}$ up-patches $C_{ijk}$.  Hence $\optup_{ijk} \le \|N'\cap
  C_{ijk}\|$, where $\optup_{ijk}$ denotes the cost of a minimum
  up-patching of $C_{ijk}$.  The problem of optimally up-patching
  $C_{ijk}$ is just an instance $I_{ijk}$ of 3D-RSA in which all
  terminals in $C_{ijk}$ have to be connected by an M-path
  to~$\hat{p}_{ijk}$.  Applying the factor-$\rho$ approximation algorithm for
  3D-RSA to each instance $I_{ijk}$ with $C_{ijk}$ relevant, we patch
  at total cost at most
  \begin{displaymath}
    \rho\sum_{C_{ijk}\text{
	relevant}}\optup_{ijk} \quad\le\quad
    \rho\sum_{C_{ijk}\text{ relevant}}\|N'\cap
    C_{ijk}\| \quad\le\quad 4\rho \|N'\|\,. 
  \end{displaymath}
  The last inequality follows from the fact that each edge of $N'$
  occurs in at most \emph{four} cuboids.  The lemma follows since
  $\|N'\|\leq 3(c^2+c)\opt$.
\end{pf}

\paragraph{Running time.}

Finally, we analyze the running time.  Let $T(n)$ denote the running
time of the algorithm applied to a set of $n$ 
terminals. The running time is dominated by patching and the recursive
slab treatment.  Using the DST algorithm of Charikar
\etal~\cite{cccdggl-aadsp-98}, patching cuboid~$C_i$ requires time  
$n_i^{O(1/\eps)}$, where $n_i$ is the number of terminals in~$C_i$.  As
each cuboid is patched at most twice and there are $c^3$ cuboids,
patching takes $O(c^3) n^{O(1/\eps)} =
n^{O(1/\eps)}$ time.  The algorithm is applied recursively to $3c$
slabs.  This yields the recurrence $T(n) = 3c T(n/c) + n^{O(1/\eps)}$,
which leads to the claimed running time.

This completes the proof of Theorem~\ref{thm:3dmain}.

\section{Open Problems}

We have presented, for any $\eps>0$, a grid-based
$O(n^\eps)$-approximation algorithm 
for $d$D-MMN.  This is a significant improvement over the ratio of
$O(n^{4/5+\eps})$ which is achieved by reducing the problem to
DSF. %
For 3D, we have described a $4(k-1)$-approximation algorithm for the case when
the terminals lie on $k \ge 2$ horizontal planes. This outperforms our
grid-based algorithm when $k \in o(n^\eps)$. 
Whereas 2D-MMN admits a 2-approximation
\cite{cnv-raamm-08,gsz-yaa2a-08,n-eprmc-05}, it remains open whether
$O(1)$- or $O(\log n)$-appro\-xi\-ma\-tion algorithms exist for higher
dimensions.

Our $O(n^{\eps})$-approximation algorithm for $d$D-MMN solves
instances of $d$D-RSA for the subproblem of patching.  We conjecture that
$d$D-RSA admits better approximation ratios.  While this is an interesting open
question, a positive result would still not be enough to improve our
approximation ratio, which is dominated by the cost of finding M-paths
inside slabs.

The complexity of the \emph{undirectional} bichromatic rectangle
piercing problem (see Section~\ref{sec:k-planes}) is still unknown.
Currently, the best 
approximation has a ratio of~4, which is (trivially) 
implied by the
result of Soto and Telha~\cite{st-2dorg-11}.  Any progress would
immediately improve the approximation ratio of our algorithm for the
$k$-plane case of 3D-MMN (for any $k>2$).

\paragraph*{Acknowledgments.}
This work was started at the 2009 Bertinoro
Workshop on Graph Drawing.  We thank the organizers Beppe Liotta and
Walter Didimo for creating an inspiring atmosphere.  We also thank
Steve Wismath, Henk Meijer, Jan Kratochv{\'i}l, and Pankaj Agarwal for
discussions.  We are indebted to Stefan Felsner for pointing us to
Soto and Telha's work~\cite{st-2dorg-11}.

\clearpage

\appendix

\section*{\Large Appendix: Extension to Higher Dimensions} 
\label{sec:dd}

We now describe the approximation algorithm for $d$D-MMN,
for $d>3$, as a generalization of the 3D-MMN idea from 
Section~\ref{sec:general-case}. 

\begin{theorem}
  For any fixed dimension $d$ and for any $\eps>0$, there
  exists an $O(n^\eps)$-approximation algorithm for $d$D-MMN.
\end{theorem}

Large parts of the algorithm and the analysis are 
straightforward generalizations of the algorithm for 3D-MMN.  The
presentation of both follows closely the 3D case. However,
Lemma~\ref{lem:dd-patching-cost}, where the cost of the 
patching procedure is bounded, requires non-trivial additional insights.

As in the 3D case we decompose the overall problem into a constant
number of instances of the directional subproblem.  The directional
subproblem consists in M-connecting any terminal pair $(t,t')$ such
that $x^i(t)\leq x^i(t')$ for any $i\in\{1,\dots,d\}$.  Here, we use
$x^i(p)$ to denote the $i$-th coordinate of point $p\in\mathbb{R}^d$.
We can decompose the general problem into $2^{d-1}$ directional
subproblems. Once again, we assume that the terminals are in general position.

\subsection*{The $d$D Grid Algorithm} 

We begin the description with a high-level summary.  To solve the
directional $d$D-MMN problem we place a $d$D-grid which partitions the
instance into 
cuboids and slabs.  Terminal pairs lying in different slabs are
handled by M-connecting each terminal to the corner of its cuboid and
then using the edges of grid. Terminal pairs from the same slab are
M-connected by applying the algorithm recursively to all slabs.  Each
slab contains only a constant fraction of the terminals.

\noindent
\smallskip
\noindent {\em Step 1: Partitioning into cuboids and slabs.} Consider
the bounding cuboid $C$ for the set $T$ of terminals and choose a
large constant $c=d^{1/\eps}$.  For each dimension $i\in\{1,\dots,d\}$
we choose $c+1$ separating planes determined by values
$x_1^i<\dots<x_{c+1}^i$.  Planes $x_1^i$ and $x_{c+1}^i$ coincide with
the boundary of $C$ in dimension $i$.  The separating planes for
dimension $i$ partition $C$ into $c$ slabs $C_j^i$, where
$j\in\{1,\dots,c\}$.  Slab $C_j^i$ is the set of points $p\in C$ such
that $x_i^j\leq x_i(p)\leq x_i^{j+1}$. We place the separating planes
so that each slab contains at most $n/c$ terminals.  Altogether we
have $d(c+1)$ separating planes.

Let $j_1,\dots,j_d\in\{1,\dots,c\}$.  The \emph{subcuboids}
$C_{j_1,\dots,j_d}$ is the set of points $p\in C$ such that
$x_{j_i}^i\leq x^i(p)\leq x_{j_i+1}^i$ for each $i\in \{1,\dots,d\}$.

Consider the $i^{th}$ dimension, $i\in\{1,\dots,d\}$ and integers
$j_k\in\{2,\dots,c\}$ for each $k\in\{1,\dots,d\}-\{i\}$.  Let $s$ be
the axis-parallel line segment that contains all points $p\in C$ such
that $x^k(p)=x_{j_k}$ for each $k\neq i$.  We call $s$ a \emph{grid
  segment for dimension~$i$}.  The \emph{grid $\cal G$} is the set of
all grid segments and there are $d(c-1)^{d-1}$ grid segments in $\cal
G$.

Given $j_k\in\{2,\dots,c\}$ for each $k\in\{1,\dots,d\}$ we call
$(x_{j_1}^1,\dots,x_{j_d}^d)$ a \emph{grid point} of $\mathcal G$ and 
there are $(c-1)^d$ grid points in total.

\smallskip
\noindent
{\em Step 2: Add M-paths between different slabs.} 
Consider two cuboids $C_{j_1,\dots,j_d}$ and $C_{j_1',\dots,j_d'}$
with $j_i<j_i'$ for each $i\in\{1,\dots,d\}$.  Any pair of terminals
$t\in C_{j_1',\dots,j_d'} $ and $t'\in C_{j_1',\dots,j_d'}$ can be
M-connected using the segments of $\mathcal G$ as long as $t$ and $t'$
are suitably connected to the corners (grid points) of their cuboids.
We use {\em patching} (described below) to connect all terminal to the
corners of their cuboid.

\smallskip {\em Patching:} Call a cuboid $C_{j_1,\dots,j_d}$
\emph{relevant} if there is a cuboid $C_{j_1',\dots,j_d'}$ that
contains at least one terminal and satisfies $j_i<j_i'$ for each
$i\in\{1,\dots,d\}$. For each relevant cuboid $C_{j_1,\dots,j_d}$, let
$p_{j_1,\dots,j_d}$ denote the grid point
$(x_{j_1+1}^i,\dots,x_{j_d+1}^d)$.  \emph{Up-patching}
$C_{j_1,\dots,j_d}$ means to M-connect every terminal in $t\in
C_{j_1,\dots,j_d}$ to $p_{j_1,\dots,j_d}$. Up-patch
$C_{j_1,\dots,j_d}$ by solving the $d$D-RSA problem with the terminals
inside $C_{j_1,\dots,j_d}$ as the terminals and $p_{j_1,\dots,j_d}$ as
the origin.

{\em Down-patching} is defined analogously; cuboid
$C_{j_1,\dots,j_d}$ is relevant if there is a non-empty cuboid
$C_{j_1',\dots,j_d'}$ with $j_i'<j_i$, $i=1,\dots,d$ and using grid
point $p_{j_1,\dots,j_d}':=(x_{j_1+1}^i,\dots,x_{j_d+1}^d)$ as origin
instead of $p_{j_1,\dots,j_d}$.

The output of this step is the union of grid $\mathcal G$ with a
network that up-patches and down-patches all relevant cuboids. This
produces M-paths between all terminal pairs in different slabs.

\smallskip
\noindent {\em Step 3: Add M-paths within slabs.}  To also connect
terminal pairs that lie in a common slab we apply the algorithm (Steps
1--3) recursively to each slab $C_j^i$ with $i\in\{1,\dots,d\}$ and
any $j\in\{1,\dots,c\}$.

\subsection*{Analysis}

We now show that the algorithm presented above
yields a feasible solution to directional
$d$D-MMN, with cost at most $O(n^\eps)\opt$, for any $\eps>0$. 
Here, \opt denotes
the cost of an optimum solution to the \emph{general} $d$D-MMN
instance rather than the minimum cost $\opt'$ achievable for the
directional subproblem.  The reason is that the cost of the grid $\cal
G$ is generally not related to $\opt'$ but to \opt.
We finish the section by arguing that the running time of the
algorithm is $n^{O(1/\eps)}$. 

\begin{lemma}[Feasibility]
  \label{lem:dd-mpaths}
  The dD grid algorithm M-connects all relevant terminal pairs. 
\end{lemma}

\begin{pf} 
  Let $(t,t')$ be a relevant terminal pair.  First suppose that $t\in
  C_{j_1,\dots,j_d}$ and $t'\in C_{j_1',\dots,j_d'}$ where $j_i<j_i'$
  for all $i\in\{1,\dots,d\}$.  Hence, $C_{j_1,\dots,j_d}$ and
  $C_{j_1',\dots,j_d'}$ are relevant for up-patching and
  down-patching, respectively.  Consider the corners
  $p_{j_1,\dots,j_d}$ of $C_{j_1,\dots,j_d}$ and the corner
  $p_{j_1',\dots,j_d'}'$ of $C_{j_1',\dots,j_d'}$. In the up-patching
  step of our algorithm we solve an RSA problem with terminals of
  $C_{j_1,\dots,j_d}$ as the input points, and corner
  $p_{j_1,\dots,j_d}$ as the origin. By definition, an RSA solution
  M-connects $t$ to $p_{j_1,\dots,j_d}$.  Similarly, down-patching
  M-connects $t'$ to $p_{j_1',\dots,j_d'}'$.  It follows that $t$ and
  $t'$ are connected, since $p_{j_1,\dots,j_d}$ and
  $p_{j_1',\dots,j_d'}'$ are M-connected via grid $\mathcal G$.  Both
  terminals are even M-connected since additionally $t\leq
  p_{j_1,\dots,j_d}\leq p_{j_1',\dots,j_d'}'\leq t'$, where $\leq$
  denotes the domination relation between points.

  Now suppose $t$ and $t'$ lie in the same slab. As the algorithm is
  applied recursively to each slab there will be a recursion step
  where $t$ and $t'$ will lie in cuboids in different slabs. Here, we
  need our assumption of general position. Applying the argument above
  to that particular recursive step completes the proof.
\end{pf}

\paragraph{Approximation ratio.}

Let $r(n)$ denote the approximation ratio of our algorithm where $n$
is the number of terminals in $T$.  The total cost of our solution
consists of the cost for the grid $\mathcal G$, the cost of
up-patching and down-patching all relevant cuboids, and the cost for
the recursive treatment of the slabs in all $d$ dimensions.  We
analyze each of these costs separately.

The grid $\mathcal G$ consists of the $d(c-1)^{d-1}$ grid segments.  The
length of any grid segment $s$ is a lower bound on $\opt$.  This holds
because there are two terminals on the boundary of $C$ whose
$L_1$-distance is at least the length of $s$.
It follows that the cost of the grid is bounded by $d(c-1)^{d-1}\opt$.

Let $r_{\text{patch}}(n)$ denote the cost of patching all relevant
cuboids as is done in Step 2. Lemma~\ref{lem:dd-dstpatch} (given
below) proves that $r_{\text{patch}}(n) = O(n^{\eps})\opt$.

Now consider the recursive application of the algorithm to all slabs
$C_j^i$, where $i\in\{1,\dots,d\}$ and $j\in\{1,\dots,c\}$.  First
recall that we placed the separating planes so that $|C_j^i|\leq n/c$
for any $i\in\{1,\dots,d\}$ and any $j\in\{1,\dots,c\}$.

Consider dimension $i\in\{1,\dots,d\}$.  Let $\opt_j^i$ be the optimum
cost for M-connecting \emph{all} (not only relevant) terminal pairs in
slab $C_j^i$, where $j\in\{1,\dots,c\}$.  Slightly abusing notation, we
write $\Nopt\cap C_j^i$ for the set of line segments of $\Nopt$ that
are completely contained in the slab $C_j^i$.  Observe that $\Nopt\cap
C_j^i$, forms a feasible solution for $C_j^i$. Thus $\opt_j^i\leq
\|\Nopt\cap C_j^i\|$.  Each such $C_j^i$ contains at most $n/c$
terminals, and therefore the total cost of the solutions for the all
slabs $C_j^i$ of dimension $i$ is at most 
\begin{displaymath}
  \sum_{j=1}^cr\left(\frac{n}{c}\right)\opt_j^i\leq
  r\left(\frac{n}{c}\right)\sum_{j=1}^c\|\Nopt\cap C_j^i\|\leq
  r\left(\frac{n}{c}\right)\opt\,.
\end{displaymath}
Summing all costs, we obtain the following recursive equation for $r(n)$
\begin{displaymath}
  r(n)\opt \leq dc^{d-1}\cdot\opt+d\cdot
  r\left(\frac{n}{c}\right)\opt+r_{\text{patch}}(n)\opt\,. 
\end{displaymath}
Hence $r(n)=O(n^{\max\{\eps,\log_cd\}})$.  Choosing
$c \ge d^{1/\eps}$, as in Step 1, yields $O(n^\eps)$ proving the
approximation ratio claimed in Theorem~\ref{thm:3dmain}.

\begin{lemma}
  \label{lem:dd-dstpatch} 
  The cost of patching all relevant cuboids, $\rpatch(n)$, is
  $O(n^{\eps})\opt$.
\end{lemma}

\begin{pf}
  First consider up-patching. Lemma~\ref{lem:dd-patching-cost} (below)
  shows the existence of a near optimal network that up-patches all
  relevant cuboids. Lemma~\ref{lem:dd-patching-algorithm} shows that
  by reducing the patching problem to $d$D-RSA, we can find such a
  network of cost $O(\rho)\opt$, where $\rho$ is the approximation
  factor of $d$D-RSA.

  Analogously to the 3D-case there is a approximation-preserving
  reduction from $d$D-RSA to DST (see
  Section~\ref{sec:relat-stein-type}), which implies that $d$D-RSA is
  approximable within a factor $O(n^\eps)$ for any $\eps>0$.  Hence
  the same approximation factor can be achieved for $d$D-RSA by
  choosing $\epsilon$ sufficiently small.

  The lemma follows as the analysis holds analogously for
  down-patching.
\end{pf}

\begin{lemma}
  \label{lem:dd-patching-cost}
  There exists an M-network of total cost at most $(c+1)^d\opt$
  that up-patches all relevant cuboids.
\end{lemma}

\begin{pf}
  Let $I\subseteq\{1,\dots,d\}$ be a set of dimensions.  For every
  $i\in I$  we choose a separating plane $x_{j_i}^i$ where
  $j_i\in\{1,\dots,c\}$.  Let $J$ be the set of 
  these separating planes and let $C(J)$ be the intersection of $C$ with
  all separating planes in $J$.  We call $C(J)$ a \emph{$C$-face}.
  There are most $(c+1)^d$ such $C$-faces.  Project $\Nopt$ onto each
  $C$-face.  Let $N'$ be the union of all these
  projections.  Clearly, the cost of $N'$ is at most $(c+1)^d\opt$.

  We claim that $N'$ up-patches all relevant cuboids.  To this end,
  let $(t,t')$ be a relevant terminal pair such that $t\in
  C_{j_1,\dots,j_d}$, $t'\in C_{j_1',\dots,j_d'}$ and $j_i<j_i'$ for
  all $i\in\{1,\dots,d\}$.  

  We claim that there is an M-path $\pi_t$ from $t$ to
  $p_{j_1,\dots,j_d}$ in $N'$.  To see this, traverse the M-path $\pi_{tt'}$ in
  $\Nopt$ connecting $t$ and $t'$, starting from $t$.  We assume
  w.l.o.g.\ that the separating planes that bound cuboid
  $C_{j_1,\dots,j_d}$ are entered by $\pi_{tt'}$ in the order
  $(x_{j_1+1}^1,\dots,x_{j_d+1}^d)$.  The desired path $\pi_t$ starts
  at $t$ and follows $\pi$ until the separating plane $x_{j_1+1}^1$ is
  entered.  From this point on we follow the projection
  $\pi_{tt'}(x_{j_1+1}^1)$ of $\pi_{tt'}$ onto $C$-face
  $C(x_{j_1+1}^1)$.  If $\pi_{tt'}(x_{j_1+1}^1)$ enters $x_{j_2+1}^2$
  we follow the projection $\pi_{tt'}(x_{j_1+1}^1,x_{j_2+1}^2)$ of
  $\pi_{tt'}$ onto $C(x_{j_1+1}^1,x_{j_2+1}^2))$.  We proceed in this
  fashion until we reach the $C$-face 
  $C(x_{j_1+1}^1,\dots,x_{j_d+1}^d)$, which is just the corner
  $p_{j_1,\dots,j_d}$.  Since $N'$ contains the projection of
  $\pi_{tt'}$ onto each $C$-face, the path $\pi_t$ described above is
  contained in $N'$.  This reasoning remains valid if the separating
  planes that bound $C_{j_1,\dots,j_d}$ are entered in an arbitrary
  order, as we projected $\Nopt$ onto each $C$-face.
\end{pf}

\begin{lemma}
  \label{lem:dd-patching-algorithm}
  Given an efficient algorithm that approximates $d$D-RSA within a
  factor of $\rho$, we can efficiently up-patch all relevant cuboids
  at cost at most $(2(c+1))^d \rho\opt$.
\end{lemma}

\begin{pf}
  In Lemma~\ref{lem:dd-patching-cost}, we showed the existence of a
  network $N'$ that up-patches all relevant cuboids at low cost.  Now
  consider an arbitrary relevant cuboid $C_{j_1,\dots,j_d}$.  Clearly
  $N'\cap C_{j_1,\dots,j_d}$ up-patches $C_{j_1,\dots,j_d}$.  Hence
  $\optup_{j_1,\dots,j_d} \le \|N'\cap C_{j_1,\dots,j_d}\|$, where
  $\optup_{j_1,\dots,j_d}$ denotes the cost of a minimum up-patching
  of $C_{j_1,\dots,j_d}$.  The problem of optimally up-patching
  $C_{j_1,\dots,j_d}$ is just an instance $I_{j_1,\dots,j_d}$ of
  $d$D-RSA, in which all terminals in $C_{j_1,\dots,j_d}$ have to be
  connected by an M-path to $p_{j_1,\dots,j_d}$.  Applying the
  factor-$\rho$ approximation algorithm for $d$D-RSA to each instance
  $I_{j_1,\dots,j_d}$ with $C_{j_1,\dots,j_d}$ relevant, we patch at
  total cost at most
  \begin{displaymath}
    \rho\sum_{C_{j_1,\dots,j_d}\text{
	relevant}}\optup_{j_1,\dots,j_d} \quad\le\quad
    \rho\sum_{C_{j_1,\dots,j_d}\text{ relevant}}\|N'\cap
    C_{j_1,\dots,j_d}\| \quad\le\quad 2^d\rho \|N'\|\,. 
  \end{displaymath}
  The last inequality follows from the fact that each segment of $N'$
  occurs in at most $2^d$ cuboids.  The lemma follows since
  $\|N'\|\leq (c+1)^d\opt$.
\end{pf}

\paragraph{Running time.}

Let $T(n)$ denote the running time of the algorithm for $n$
terminals. The running time is dominated by patching and the recursive
treatment of slabs. Using the DST algorithm of Charikar
\etal~\cite{cccdggl-aadsp-98}, patching cuboid $C_j^i$ requires time 
$(n_j^i)^{O(1/\eps)}$, where $n_j^i$ is the number of terminals in
$C_j^i$. As each cuboid is patched at most twice and there are $c^d$
cuboids, patching requires total time $O(c^d) n^{O(1/\eps)} =
n^{O(1/\eps)}$. The algorithm is applied recursively to $dc$
slabs. This yields the recurrence $T(n) = dc T(n/c) + n^{O(1/\eps)}$,
which leads to the claimed running time.


\begin{thebibliography}{10}
\providecommand{\url}[1]{{#1}}
\providecommand{\urlprefix}{URL }
\expandafter\ifx\csname urlstyle\endcsname\relax
  \providecommand{\doi}[1]{DOI~\discretionary{}{}{}#1}\else
  \providecommand{\doi}{DOI~\discretionary{}{}{}\begingroup
  \urlstyle{rm}\Url}\fi

\bibitem{a-asnph-03}
Arora, S.: Approximation schemes for {NP}-hard geometric optimization problems:
  A survey.
\newblock Math. Program. \textbf{97}(1--2), 43--69 (2003)

\bibitem{admss-esstl-95}
Arya, S., Das, G., Mount, D.M., Salowe, J.S., Smid, M.: {Euclidean} spanners:
  Short, thin, and lanky.
\newblock In: Proc. 27th Annu. ACM Sympos. Theory Comput. (STOC'95), pp.
  489--498 (1995).
\newblock \urlprefix\url{http://dx.doi.org/10.1145/225058.225191}

\bibitem{bwws-mmnpa-06}
Benkert, M., Wolff, A., Widmann, F., Shirabe, T.: The minimum {Manhattan}
  network problem: Approximations and exact solutions.
\newblock Comput. Geom. Theory Appl. \textbf{35}(3), 188--208 (2006).
\newblock \urlprefix\url{http://dx.doi.org/10.1016/j.comgeo.2005.09.004}

\bibitem{cccdggl-aadsp-98}
Charikar, M., Chekuri, C., Cheung, T.Y., Dai, Z., Goel, A., Guha, S., Li, M.:
  Approximation algorithms for directed {Steiner} problems.
\newblock In: Proc. 9th ACM-SIAM Sympos. Discrete Algorithms (SODA'98), pp.
  192--200 (1998)

\bibitem{cnv-raamm-08}
Chepoi, V., Nouioua, K., Vax\`{e}s, Y.: A rounding algorithm for approximating
  minimum {Manhattan} networks.
\newblock Theor. Comput. Sci. \textbf{390}(1), 56--69 (2008).
\newblock \urlprefix\url{http://dx.doi.org/10.1016/j.tcs.2007.10.013}

\bibitem{cgs-mmnnp-11}
Chin, F., Guo, Z., Sun, H.: Minimum {Manhattan} network is {NP}-complete.
\newblock Discrete Comput. Geom. \textbf{45}, 701--722 (2011).
\newblock \urlprefix\url{http://dx.doi.org/10.1007/s00454-011-9342-z}

\bibitem{fkn-iaadsf-09}
Feldman, M., Kortsarz, G., Nutov, Z.: Improved approximating algorithms for
  directed {Steiner} forest.
\newblock In: Proc. 20th ACM-SIAM Sympos. Discrete Algorithms (SODA'09), pp.
  922--931 (2009)

\bibitem{fs-s3amm-08t}
Fuchs, B., Schulze, A.: A simple 3-approximation of minimum {Manhattan}
  networks.
\newblock Tech. Rep. 570, Zentrum f{\"u}r Angewandte Informatik K{\"o}ln
  (2008).
\newblock \urlprefix\url{http://e-archive.informatik.uni-koeln.de/570}

\bibitem{gln-ammn-01}
Gudmundsson, J., Levcopoulos, C., Narasimhan, G.: Approximating a minimum
  {Manhattan} network.
\newblock Nordic J. Comput. \textbf{8}, 219--232 (2001)

\bibitem{gsz-yaa2a-08}
Guo, Z., Sun, H., Zhu, H.: Greedy construction of 2-approximation minimum
  {Manhattan} network.
\newblock In: S.~Hong, H.~Nagamochi, T.~Fukunaga (eds.) Proc. 19th Annu.
  Internat. Sympos. Algorithms Comput. (ISAAC'08), \emph{Lecture Notes Comput.
  Sci.}, vol. 5369, pp. 4--15. Springer, Berlin (2008).
\newblock \urlprefix\url{http://dx.doi.org/10.1007/978-3-540-92182-0_4}

\bibitem{kia-iammn-02}
Kato, R., Imai, K., Asano, T.: An improved algorithm for the minimum
  {Manhattan} network problem.
\newblock In: P.~Bose, P.~Morin (eds.) Proc. 13th Annu. Internat. Sympos.
  Algorithms Comput. (ISAAC'02), \emph{Lecture Notes Comput. Sci.}, vol. 2518,
  pp. 344--356. Springer, Berlin (2002).
\newblock
  \urlprefix\url{http://link.springer.de/link/service/series/0558/bibs/2518/25%
180344.htm}

\bibitem{lap-pafst-03}
Lam, F., Alexandersson, M., Pachter, L.: Picking alignments from ({Steiner})
  trees.
\newblock J. Comput. Biol. \textbf{10}, 509--520 (2003)

\bibitem{lr-ptasrsap-00}
Lu, B., Ruan, L.: Polynomial time approximation scheme for the rectilinear
  {Steiner} arborescence problem.
\newblock J. Comb. Optim. \textbf{4}(3), 357--363 (2000)

\bibitem{msu-mmnp3-09}
Mu{\~n}oz, X., Seibert, S., Unger, W.: The minimal {Manhattan} network problem
  in three dimensions.
\newblock In: S.~Das, R.~Uehara (eds.) Proc. 3rd Internat. Workshop Algorithms
  Comput. (WALCOM'09), \emph{Lecture Notes Comput. Sci.}, vol. 5431, pp.
  369--380. Springer, Berlin (2009).
\newblock \urlprefix\url{http://dx.doi.org/10.1007/978-3-642-00202-1_32}

\bibitem{n-eprmc-05}
Nouioua, K.: Enveloppes de {Pareto} et r{\'e}seaux de {Manhattan}:
  Caract{\'e}risations et algorithmes.
\newblock Ph.D. thesis, Universit{\'e} de la M{\'e}diterran{\'e}e (2005).
\newblock
  \urlprefix\url{http://www.lif-sud.univ-mrs.fr/~karim/download/THESE_NOUIOUA.%
pdf}

\bibitem{su-15amm-05}
Seibert, S., Unger, W.: A 1.5-approximation of the minimal {Manhattan} network
  problem.
\newblock In: X.~Deng, D.~Du (eds.) Proc. 16th Intern. Symp. Algorithms Comput.
  (ISAAC'05), \emph{Lecture Notes Comput. Sci.}, vol. 3827, pp. 246--255.
  Springer, Berlin (2005).
\newblock \urlprefix\url{http://dx.doi.org/10.1007/11602613_26}

\bibitem{st-2dorg-11}
Soto, J.A., Telha, C.: Jump number of two-directional orthogonal ray graphs.
\newblock In: O.~G{\"u}nl{\"u}k, G.~Woeginger (eds.) Proc. 16th Internat. Conf.
  Integer Prog. Comb. Optimization (IPCO'11), \emph{Lecture Notes Comput.
  Sci.}, vol. 6655, pp. 389--403. Springer, Berlin (2011).
\newblock \urlprefix\url{http://dx.doi.org/10.1007/978-3-642-20807-2_31}

\bibitem{zelikovsky-adstapprox-97}
Zelikovsky, A.: A series of approximation algorithms for the acyclic directed
  {Steiner} tree problem.
\newblock Algorithmica \textbf{18}(1), 99--110 (1997)

\end{thebibliography}
\end{document}